\documentclass[aps,twocolumn,twoside,floatfix,nofootinbib,a4paper]{revtex4}
\usepackage{graphicx}
\usepackage{marginnote}

\usepackage{hyperref}
\usepackage{enumitem,kantlipsum}
\usepackage[usenames,dvipsnames]{color}
\usepackage{amsmath}
\usepackage{amssymb}
\usepackage{amsfonts}
\usepackage{mathrsfs}
\usepackage{bm}
\usepackage[all]{xy}
\usepackage{hyperref}
\usepackage{tikz-cd}

\newcommand{\beq}{\begin{equation}}
\newcommand{\eeq}{\end{equation}}
\newcommand{\bea}{\begin{eqnarray}}
\newcommand{\eea}{\end{eqnarray}}
	
\newcommand{\ea}{\end{align*}}

\newcommand{\bma}{\begin{pmatrix}}
\newcommand{\ema}{\end{pmatrix}}

\usepackage[percent]{overpic}
\usepackage{overpic}

\newcommand{\Msun}{\text{M}_{\odot}}

\newcommand{\appropto}{\mathrel{\vcenter{
  \offinterlineskip\halign{\hfil$##$\cr
    \propto\cr\noalign{\kern2pt}\sim\cr\noalign{\kern-2pt}}}}}

\begin{document}
\title{Final parsec evolution in the presence of intermediate mass black holes}
\author{Fan Zhang} 
\affiliation{Gravitational wave and Cosmology Laboratory, Department of Astronomy, Beijing Normal University, Beijing 100875, China}
\affiliation{Advanced Institute of Natural Sciences, Beijing Normal University at Zhuhai 519087, China}

\date{\today}

\begin{abstract}
\begin{center}
\begin{minipage}[c]{0.9\textwidth}
In this short note, we draw attention to the possibility that, under favorable conditions, the final parsec problem could be alleviated by the presence of a moderate population of intermediate mass black holes in the centers of merged galaxies.  
\end{minipage} 
 \end{center}
  \end{abstract}
\maketitle

\raggedbottom

\section{Introduction \label{sec:Intro}}
Recent observations have suggested that supermassive black hole (SMBH) binaries should reach merger well within a Hubble time. This runs counter to those theoretical computations predicting that after dynamical friction loses effectiveness, but prior to gravitational wave (GW) emission takes over, beginning at around one parsec separation, the binary hardening would stall into a torturously slow pace. Such stalling would prevent merger events to ever have happened given the age of our Universe, contradicting observational evidences, thus giving rise to the so-called ``final parsec problem''. 

Specifically, the GW amplitude inferred by pulsar timing array observations \cite{2023ApJ...951L...8A,2023ApJ...951L...6R,2023A&A...678A..50E,2023RAA....23g5024X} is achievable with a short binary lifespan \cite{Agazie_2023}. There is, furthermore, the discovery \cite{2023ApJ...957..107W} of a SMBH binary system just about to enter the GW dominated regime, signifying that the final parsec does not take longer than a Hubble time. Moreover, the scarcity of direct observational evidences for SMBH binaries with roughly a parsec separation, despite extensive surveys looking for them, could be explained if the binaries do not persist for a long time in a stalled state \cite{2011MNRAS.410.2113B}. The observational evidences for a diminishment of stalling is mounting, so we would have to look carefully at theoretical predictions for omissions. 

The most important ingredient when it comes to impacting the evolution of SMBH binaries is their surrounding gravitational environment. The stellar cluster enclosing a SMBH in either of the pre-merger galaxies contains diverse populations, as should the merged core hosting the SMBH binary. Historically, literature has modeled the steady state number density of main sequence stars, white dwarfs, neutron stars, and stellar mass black holes (sBHs), seeing them in rough total population ratios of $1:0.1:0.01:0.001$ \cite{2023MNRAS.524.2033S}, and spatially distributed according to inverse power laws with indices $\alpha \sim 1.4-2.0$, with the heavier population more concentrated towards the center of gravity due to mass segregation \cite{2006ApJ...645L.133H}.  

However, as will be discussed in Sec.~\ref{sec:IMBHpop} below, there could well be a non-negligible population of intermediate mass black holes (IMBHs) lurking in the galactic centers as well, possibly even incidentally conforming to the geometric sequence for the abundances of the last paragraph by filling in the next slot at $0.0001$ times the main sequence stellar population (note, this optimistic scenario is not far off from the upper limit due to dark matter considerations discussed in Appendix \ref{sec:DM}). These IMBHs have much greater masses, and therefore could act quite differently to the already accounted-for species mentioned above. Subsequently, a proper acknowledgement of their presence may well yield more substantial alterations to the hardening processes of the SMBH binaries, possibly to the desirable effect of alleviating the final parsec problem.  

Immediately, we notice that IMBHs would concentrate in an even steeper cuspy number density profile towards the SMBH binary than even sBHs, due to their much greater masses. For intuition, imagine the extreme limit where the IMBHs are just a little lighter than the SMBHs themselves, then they should essentially be at where the SMBHs are, as they all sink via the same dynamical friction process. To see the consequence of this crowding, note first that the radius of influence for each of the SMBH is \cite{2023MNRAS.524.2033S} 
\bea
r_{\rm inf} \simeq 2.4{\rm pc}\sqrt{\frac{M_{\rm SMBH}}{4\times 10^6 \Msun}}\,,
\eea
which is on the order of tens of parsecs for a say, $10^8\Msun$ SMBH, much larger than the stalling-epoch parsec separation between the SMBHs. If we regard all IMBHs within $\sim r_{\rm inf}$ to the center of the binary as being eligible for being scattered by it \cite{2006ApJ...651..392S}, then with the aforementioned power law cuspy number density profile, it is easy to show that\footnote{After regularizing by cutting off the profile at roughly the binary separation, which is much larger than the cutoff given in \cite{2023MNRAS.524.2033S} for single SMBH. The numbers are more conservative for larger cutoffs.} $90\%$ of all the IMBHs in the merged core will be available for scattering if $\alpha=2$ as with sBHs. However, $\alpha$ should be larger for heavier holes, giving us $99\%$ availability when $\alpha = 3$, and another $9$ added at the tail of the decimals whenever $\alpha$ goes up by one. 
In other words, even an overall relative dearth of IMBHs in the core of the merged galaxies would not preclude them from being relatively more abundant near the SMBHs, where as far as resolving the final parsec problem is concerned, is the vital playground where the relevant slingshot mechanism (see Sec.~\ref{sec:finalpc} below) operates.

More importantly, the greater mass means whenever an IMBH gets an energy boost from the binary via slingshot interactions with the SMBHs, it takes away far greater energy than a regular stellar mass object would. For intuition, imagine almost headon slingshot interactions as being analogues to scatterings between a billiard ball and a marble, the billiard ball would be affected much more severely if the marble is of a greater mass. Because the binaries harden via slingshotting (traditionally) stars, it stands to argue that IMBHs would present interesting alternative scattering agents that may well substantially hasten the final parsec separation evolution of the SMBH binaries. In Sec.~\ref{sec:finalpc}, we review in slightly more details the binary hardening process, highlighting where the presence IMBHs could make a difference. Then in Sec.~\ref{sec:IMBHpop}, we make one (but not unique) argument for why there could indeed be IMBHs present where needed. Subsequently, in Sec.~\ref{sec:solvepc}, we estimate how much of a difference these IMBHs could make to the final parsec problem, before finally discussing how to verify or falsify observationally that such a IMBH population actually do exist in Sec.~\ref{sec:discussion}.

\section{Final parsec problem \label{sec:finalpc}}
For a quick overview of the final parsec problem, we begin by recalling the SMBH inspiral stages outlined in \cite{1980Natur.287..307B} (see also \cite{1997MNRAS.284..318S,2003ApJ...596..860M,2003CQGra..20S..45S,2004ApJ...602...93M}). Note we will inherit their terminology of referring to the scattering agents as stars, although we will switch to IMBHs later on. 
\begin{enumerate}
\item 
When two galaxies merge, their cores merge by dynamical friction, and then the SMBHs, as well as the IMBHs, originally contained within the nuclear clusters of the respective galaxies, sink towards the center also by dynamical friction, with a timescale (defined as $|r/\dot{r}|$, with $r$ being separation between the SMBHs; this quantity gives the time it takes til merger if the current $\dot{r}$ is maintained)
\bea \label{eq:dftime}
t_{\bf df} \propto \frac{\sigma_c r_c^2 QM_{\rm SMBH}}{\ln N}\,,
\eea
where $N$ is the number of stars in the merged core, $\sigma_c$ is their dispersion velocity, $r_c$ is the core radius, while $QM_{\rm SMBH}$ is the mass of the smaller SMBH, with $M_{\rm SMBH}$ being the mass of the heavier hole and $Q$ being the mass ratio. The timescale is set by the smaller hole, because it sinks more slowly. 

\item 
After the SMBHs get close enough to become gravitational bound to each other (when mass enclosed by the orbit becomes comparable to that of the holes), the orbit keeps shrinking efficiently by dynamical friction at the same timescale $t_{\rm df} \propto r^0$. This estimate, based on dynamical friction experienced by an individual (i.e., stays the same if the other hole is absent) SMBH sinking towards the center of gravity of the merged galaxy, is valid when the SMBHs are so far separated that their interactions with the stars can essentially be seen as the stars being slingshot by individual solitary passing holes. This picture is valid until such a time that a star's encounter with the holes lasts longer than the orbital period of the binary. After this time, three body interactions between the star and both SMBHs become important. The energy transfer from the binary onto the stars being slingshot out becomes less efficient, reflected in a diminished effective impact parameter (scatterings where the holes contribute to stellar energy pickup in offsetting ways correspond to the ineffective segments of the impact parameter, and thus excluded) that picks up an additional dependence on $r$. However, this only affects the dynamical friction timescale $t_{\rm df}$ via the logarithm term, so it remains roughly constant. 

\item 
Further down the line though, as the binary becomes more compact, the SMBHs essentially become pinned down at a single spatial location (this is not entirely accurate though, the binary can wander a bit as a whole \cite{2004ApJ...602...93M}), thus can no longer simply randomly run into stars on their long tours across the cluster. Instead, for slingshot transfers of energy out of the binary to continue to work, the stars will have to fall towards the binary in such a way that their closest point of approach is sufficiently close to the center that they can tell the two members of the binary apart, or else their transits will merely be like passing close to a single gravitating mass, and their energies will be conserved (see also discussion in the previous paragraph). This means only low angular momentum stars on more elongated, almost radial, orbits, i.e., those in the so-called loss (because stars in it get flung out of the cluster by the binary) cone, are eligible to participate. 

As stars initially in the loss cones pick up energy during one or several encounters with the binary and end up getting ejected from the core, the loss cone gets depleted and needs replenishment through the two-body relaxation process between the stars. Depending on how efficiently this replenishment happens, there are two possibilities: 
\begin{enumerate} [wide, labelwidth=!, labelindent=0pt]
\item 
If the loss cone is replenished by two-body relaxation more quickly than it is cleared out (the ``pinhole'' regime), then the timescale of this stage, with a loss cone kept at full, is given by $t_{\rm sl:rep} \propto r^{-1}$, which increases as the binary hardens, but not too dramatically. This slowdown of hardening happens because as the binary continues to harden, the loss cone ``solid angle''\footnote{One can rescale the actual specific angular momentum by that at the same energy if circular orbit \cite{1978ApJ...226.1087C,1999MNRAS.309..447M}, thus yielding an angle-like dimensionless quantity that offers a measure of how close to radial the stellar orbit is (this rescaled quantity equals unity minus eccentricity squared).} \cite{1980Natur.287..307B}  declines as there needs to be a more precise aiming at the center of the binary (velocity of stars becoming more radial), for the star to be able to get inside a smaller binary region. As a result, there are simply less stars in the (kept full) loss cone that come in to drain the binary orbital energy (note also, the average fractional energy drain during each slingshot encounter is independent of the binary separation \cite{1980Natur.287..307B}).  

\item 
Alternatively, if the two-body relaxation time
\bea \label{eq:reltime}
t_{\rm rx} \sim \frac{N}{10 \ln N}\frac{r_c}{\sigma_c}
\eea
of the stellar cluster (more precisely that part of the core stellar population that is gravitationally dominated by the SMBHs, see discussions in \cite{1978ApJ...226.1087C} regarding the zero binding energy boundary) is too large, which tends to be the case for merged cores containing many stars, the depletion can be quicker than replenishment (the ``diffusion'' or thermal relaxation regime). In this case, the total capacity of how many stars can be hosted in the loss cone becomes irrelevant. Instead, as soon as a star wanders into the loss cone, it gets slingshot out, so the timescale $t_{\rm sl:dep}$ is set by the rate of such ``accidental'' events occurring (and the fractional energy loss per encounter). This rate, or replenishing flux into the loss cone, is insensitive to $r$ \cite{2002MNRAS.331..935Y} (as does the fractional energy loss), depending on it only via a logarithm (see e.g., discussions in \cite{2003ApJ...599.1129Y}), so $t_{\rm sl:dep}$ is approximately $\propto r^0$. 

Specifically, \cite{1980Natur.287..307B} gives the estimate 
\bea\label{eq:stall}
t_{\rm sl:dep} \simeq \frac{N}{10 \ln N} \left(\frac{M_{\rm SMBH}}{M_{\rm cl}}\right)^2 t_{\rm df}\,,
\eea
where $M_{\rm cl}$ is the total mass of the star cluster. The mass ratio squared is roughly the aforementioned loss cone solid angle, which when multiplied with the cluster relaxation time according to Eq.~\eqref{eq:reltime}, gives the inverse of the loss cone repopulation rate. Assuming that the relevant relaxation/recharging timescale associated with $t_{\rm df}$ for when the binary was wide, is the cluster crossing time $r_c/\sigma_c$, then one obtains Eq.~\eqref{eq:stall} as the re-adjusted binary hardening timescale, according to the changes in the character and relaxation mechanisms of the effective scattering agents (dynamical friction can also be viewed from a slingshot perspective). 

In more details, when $r$ is large, the SMBHs traverse large distances, which multiply with the scattering cross section into large volumes that contain many stars. Thus chance encounters with the SMBHs are the dominant source of stars to be scattered, and the relaxation for the distribution of these stars occurs on spatial (rather than angular momentum or energy) redistribution timescales, which is just the cluster crossing time. This is because sans the SMBH's influence, the phase space distribution of stars in the neighborhood of the would-be SMBH returns to the undisturbed state (e.g., the wake behind the SMBH dissipates, and stars repossess speeds commensurate to $\sigma_c$) after the stars in the cluster have reshuffled over the crossing timescale, so it is mostly those stars untainted by past encounters with the SMBH that now occupy the said neighborhood. In contrast, when $r$ is small, only those stars more deliberately falling towards the binary on very elliptical orbits will be able to scatter against it, and changes in the distribution of ellipticity is accomplished by migration in specific angular momentum, which is conserved if the stars do not collide with each other, so relaxation can only happen on the two-body relaxation timescale (more accurately the loss cone replenishment timescale, as a full loss cone is the ``undisturbed'' distribution of scattering stars that we are trying to relax into). 

Note though, as compared to the original formula in \cite{1980Natur.287..307B}, we have an additional slow varying (against cluster parameters) dimensionless factor of $10\ln N$ that should have been swept along when substituting in Eq.~\eqref{eq:reltime}, but seems to have been suppressed in Eqs.~2 and 3 of \cite{1980Natur.287..307B}, perhaps due to it being roughly constant. It is however useful to make it explicit when doing semi-quantitative estimates though.  
In particular, the stalling segment in Fig.~1 of \cite{1980Natur.287..307B} only goes up to $\sim \mathcal{O}(10^{11})$yr, while computation with their Eq.~3, using the assumed parameters for that figure, should have yielded $\sim \mathcal{O}(10^{13})$yr. The extra $10\ln N$ factor clarifies this apparent ``discrepancy''. 

Furthermore, we would like to draw attention to that fact that despite the nomenclature (which is borrowed from plasma physics), the ``loss cone'' is actually a hyperboloid in the velocity space \cite{1978ApJ...226.1087C}, that reduces to roughly a cylinder when we are at a part of the orbit far away from the binary. The radius of this cylinder is the maximum transverse speed (or specific angular momentum) allowed if the star is to come close enough to the binary to be flung out. Semi-quantitatively, the radius is $\sim \sqrt{M_{\rm SMBH}/M_{\rm cl}}\sim \sqrt{r_h/r_c}$ \cite{1980Natur.287..307B} ($r_h$ is the separation when the binary becomes hard) times the maximum allowed speed for bound orbits. Therefore the relative size of the loss cylinder within the phase space region for all bound orbits is $\sim M_{\rm SMBH}/M_{\rm cl}$, linear and not quadratic as in Eq.~\eqref{eq:stall}. Alternatively, in the more ``cone''-based nomenclature, the loss cone ``opening angle'' as defined by \cite{1978ApJ...226.1087C}, as the ratio between the specific angular momentum at the edge of the loss cone and that of the circular orbit at the same energy, is $\sim \sqrt{r_h/r_c}$, so the solid angle is once again $\sim (r_h/r_c)$, linear and not quadratic. In short, because the stars do not travel on straight lines, the opening size for the radial-hugging ``solid angle'' isn't quite the geometric angle one would obtain by pointing the stars directly at the binary expanse. Instead, it varies with radial speed and scales with the mass ratio $M_{\rm SMBH}/M_{\rm cl}$ differently. As a result, the square in Eq.~\eqref{eq:stall} may perhaps be superfluous\footnote{Depending on whether we had guessed correctly the rationales leading to this expression, the original short letter contains not much details. It is with this apprehension, that we keep the original power as in \cite{1980Natur.287..307B}. It is nevertheless straightforward to evaluate the changes that altering it would bring.}. Since $M_{\rm SMBH}$ isn't usually much smaller than $M_{\rm cl}$ consisting of stars, this issue doesn't matter much for previous studies (adjusting it will make $t_{\rm sl:dep}$ even longer though), but it could be of some importance if we transition into other types of scattering agents. 

For the sake of prudence then, as a comparison, we also evoke a more recent estimate for $t_{\rm sl:dep}$, as given by Eq.~(38) of \cite{2003ApJ...599.1129Y} (note this expression is roughly consistent with Fig.~1 of \cite{1980Natur.287..307B} when $m_* \sim \mathcal{O}(1)\Msun$)
\bea\label{eq:stall2}
t_{\rm sl:dep} &\approx& 6\times 10^9 {\rm yr} \frac{(1+Q)M_{\rm SMBH}} {3.5\times 10^6 \Msun}
\times \notag \\
&&\times \frac{1 \Msun}{m_*}\frac{2\times 10^{-4} {\rm yr}^{-1}}{n_{\rm diff}}\,,
\eea
which is inversely proportional to the stellar mass $m_*$, thus demonstrates more explicitly one of the motivations for us wanting to make use of IMBHs. Essentially, because the average ejection speed of stars is dependent only on the properties of the binary (the gravitational and inertial masses are equal, thus cancel out in the computation of acceleration for test masses -- self-potential of the stellar population is ignored), thus their kinetic energies are proportional to the mass of the scattering agents, and IMBHs, being much more massive than regular stars or sBHs, can potentially drain energy much more rapidly. Eq.~\eqref{eq:stall2} is more explicit in showing that $t_{\rm sl:dep}$ is set by the fractional energy drain per slingshot, and the rate of slingshot pallets being fed, the latter is given by $n_{\rm diff}$, which is the diffusion rate of stars into the loss cone (the numerical values given in the ratio are typical for stars).  It can be computed by e.g., solving a time-independent Fokker-Planck equation for the phase space stellar distribution in a spherical galaxy \cite{1978ApJ...226.1087C,1999MNRAS.309..447M}, which ends up being roughly (see \cite{1978ApJ...226.1087C} for a more precise numerical scaling law) proportional to $M^{-1}_{\rm cl}$ \cite{2003ApJ...596..860M}, so the two terms on the bottom line of Eq.~\eqref{eq:stall2} once again yield a $\appropto N$ dependence (see \cite{2004ApJ...602...93M} for confirmation by numerical experimentation).   

Because for $\mathcal{O}(1)\Msun$ stellar masses in the core, $N$ is a very large number, e.g., $\sim \mathcal{O}(10^9)$ as assumed by \cite{1980Natur.287..307B}, we end up being encumbered by a very large $t_{\rm sl:dep}$. In particular, this stage may take a time longer than the age of the Universe, giving rise to the final parsec problem.  Note however, by tightening some of the simplifications involved, such as bringing in time dependence due to e.g., the initial fast depletion of the loss cone; accounting for the secondary slingshots (see also \cite{2020EPJP..135..104Z} for some relativistic considerations on repeated slingshots) of returned stars that haven't left the galaxy altogether and not perturbed on their long journeys by complications in the broader galactic gravitational potential; and allowing Brownian motion of the binary inside the cluster, it is possible to lessen this problem \cite{2003ApJ...596..860M}. These mechanisms (and others, see e.g., \cite{1994ApJ...436..607M,1996MNRAS.278..186V,2002ApJ...567L...9A,2002ApJ...578..775B,2006ApJ...651.1059I,2006ApJ...642L..21B}) are likely at work, and the enlistment of IMBHs may provide the additional enforcement needed to help close the remaining gap for the more massive SMBH binaries. 
\end{enumerate}

\item 
Eventually, GW radiation takes over as the dominant channel of energy loss that quickly decreases binary separation as \cite{1964PhRv..136.1224P}
\bea \label{eq:GWtime}
t_{\rm gw} \propto Q^{-1} M^{-3}_{\rm SMBH} r^4 (1-e^2)^{7/2}\,,
\eea   
where $e$ is the binary eccentricity (note we have included it only to provide a convenient reference for later consideration, most discussions in this section assume circular orbits). 
\end{enumerate}
Historically, only populations with stellar masses had been considered as scattering agents for the slingshots, but IMBHs could be present as well, as we will argue in the next section. A cluster of IMBHs can have a much smaller population size, for a total cluster mass that is comparable to stellar counterparts, thus substantially decrease $t_{\rm sl:dep}$ as per Eqs.~\eqref{eq:stall} and \eqref{eq:stall2}. 

\section{IMBHs in galactic centers \label{sec:IMBHpop}}
IMBHs are expected to form in stellar clusters, including but not confined to the nuclear clusters \cite{2022ApJ...927..231F,2022ApJ...929L..22R,2023MNRAS.523.4227A,2023PhRvD.108h3012K} at galactic centers. From an abstract point of view, because the heat capacity of a self-gravitating system is negative (long range forces like gravity violate the additivity property of thermodynamics \cite{2006JPhCS..31...18B}; cf., gravothermal catastrophe \cite{1980MNRAS.191..483L}), galaxies or stellar clusters will evolve towards the most concentrated form of matter distribution, that of black holes, once their internal energy source in the form of fusion within stars begins to get exhausted and their energy leakage into the surrounding environment isn't shut off (and dark energy doesn't grow so quickly that it tears everything apart before collapses had a chance to run their courses). While galaxies form SMBHs, their miniature cousins, stellar clusters, could form IMBHs on shorter timescales, and these IMBHs should also participate in the grander scheme of the collapse of their entire host galaxies, by merging with the central SMBHs, sooner or eventually. Therefore, while the nuclear clusters are the densest, thus ideal incubators, those IMBHs born elsewhere could still make their way to the centers of galaxies \cite{2005ApJ...618..426M,2014ApJ...785...71G,2022ApJ...939...97F}. Conveniently, we may well need IMBHs to help explain the presence and properties of massive stars near the SMBH in our own Galaxy \cite{2003ApJ...593L..77H,2005ApJ...628..236G,2020ApJ...896..137G,2020ApJ...905..169Z,2020ApJ...888L...8N} (see also \cite{2020A&A...636L...5G} for observational constraint), or the hypervelocity stars \cite{2003ApJ...599.1129Y,2006MNRAS.372..174B,2008MNRAS.384..323L,2019ApJ...878...17R}. Observationally, there are also hints that our own Galactic center may indeed host IMBHs \cite{2004A&A...423..155M,2017NatAs...1..709O}.

It is therefore not outrageous to expect there to be a moderate (subdominant in total mass to stars, see Appendix \ref{sec:DM}) population of IMBHs at the very center of galaxies, close to where SMBHs reside. In fact, the merger of a cluster of IMBHs themselves may have built the SMBHs in the first place (see e.g., \cite{2001ApJ...562L..19E,2004ApJ...614..864M,2017NatAs...1..709O,2020MNRAS.496..921W}). Indeed, creating an additional channel for the growth of SMBHs apart from galactic mergers (removing the final parsec hurdle also makes viable the SMBH mergers as a growth path \cite{2007ApJ...669...67H}) could allow SMBHs to grow faster than their host galaxies, explaining the observations by e.g., \cite{2018MNRAS.475.1887Y}. For IMBH accretion to serve as a substantial growth engine for SMBHs, there needs to be a steady stream of IMBH supplied to the galactic centers and be captured by the SMBHs (see e.g., \cite{2003ApJ...596..860M,2005ApJ...618..426M} for discussions on the lack of fatal stalling for IMBH-SMBH mergers), and possibly more would remain unmerged with the SMBHs. There would thus be ($M_{\rm S/IMBH}$ are the S/IMBH masses)
\bea \label{eq:BHNum}
N_{\rm IMBH} = \frac{M_{\rm SMBH}}{ M_{\rm IMBH}} \frac{1-\xi}{\xi}
\eea
IMBHs close to an SMBH. Note there should also be an overall suppression factor as SMBH mergers also contribute to the growth of $M_{\rm SMBH}$. Because we are not sure how galactic mergers have proceeded through cosmic time, especially in light of recent JWST observations of early massive galaxies (see e.g., \cite{2023ApJ...948..126G}), we leave out this factor, noting only that it could be absorbed by $\xi$. 

The parameter $\xi\leq 1$ is the cumulative capture rate obtained by integrating a likely complicated and evolving survival probability, until a time that we are interested in, e.g., when two SMBHs are $\sim \mathcal{O}(1)$pc apart for the discussions in Sec.~\ref{sec:finalpc}.  
As an example of why survival probability would change over time, note that when the SMBH has grown to larger masses, the SMBH-IMBH bound systems would begin to suffer from the same final parsec issue as per Eq.~\ref{eq:stall2}, so survival probability of IMBHs could increase over time. Counterbalancing this effect though, is the build up of IMBH population near the SMBH over time, which induces eccentricity for each other through mutual scattering, that hasten their fate of being dynamically captured by the SMBH. 

The total mass of the cluster of IMBHs is then $\sim M_{\rm SMBH} (1-\xi)/\xi$, so would be the same order of magnitude as the SMBH mass if SMBH manages to capture around half of the IMBHs. Fig.~4 of \cite{2022ApJ...929L..22R} suggests that indeed, around two thirds of the IMBHs formed in situ in the nuclear stellar cluster, from collisions between sBHs and main sequence stars (see Appendix \ref{sec:formation} for a partial summary of other IMBH formation channels), fall into the SMBH after evolving for around $10$Gyr, roughly the age of galaxies. Note also that the sBH seeds and the stars that they consume may be continuously replenished over time in the real world, so there could also be IMBHs that have not run for the full $10$Gyr. Because formation happens prior to subsequent merger with the SMBH, a greater proportion of them would have formed but have not yet fallen into the SMBH, than those that have been simulated for the full $10$Gyr, so the actual capture rate would be somewhat lower than $2/3$. 

Moreover, of the $500$ seed sBHs (initially situated between $0.001$pc and $0.1$pc from the SMBH) simulated by \cite{2022ApJ...929L..22R}, around a third end up becoming IMBHs after evolving for a full $10$Gyr (inclusive of those that have merged with the SMBH). The present day ratio of IMBHs and sBHs would be dependent on the uncertain sBH and stellar mass replenish rates though, as some of the late-addition sBHs would not have had enough time to grow. Nevertheless, the middle panel of Fig.~4 in \cite{2022ApJ...929L..22R} shows that IMBHs have already been seen falling into the SMBH as early as $\sim 1$Gyr, and these merged population grows quickly after the Gyr mark, which is consistent with the estimate by \cite{2018MNRAS.477.4423A} that it takes several Gyr for a SMBH to swallow an IMBH. In other words, many of the IMBHs likely would have formed within a Gyr, and then hang around the SMBH for several Gyr afterwards, before getting consumed. If this is so, most of the sBHs, apart from those introduced to the galactic center within the last Gyr, would have been effective as IMBH growth seeds, so long as they are also spatially located sufficiently close to the SMBH. Indeed, because the sBHs are distributed according to a steep cuspy power law with index $\alpha = 2$ \cite{2006ApJ...645L.133H}, nearly all of the $\sim \mathcal{O}(10^4)$ sBHs in the central parsec of e.g., our own Milky Way \cite{1993ApJ...408..496M,2000ApJ...545..847M,2006ApJ...649...91F,2018Natur.556...70H} would be eligible seeds. 

More quantitatively, let $t$ be look back time, $\eta(t)$ be the rate at which sBHs are introduced into the galactic center, $\varphi(t)$ be the percentage of IMBHs that have grown from seeds introduced at time $t$ that have survived today, the the ratio $\mathcal{R}$ of IMBHs and sBHs around the SMBH today is given by (ignoring the depletion of sBHs including those that became IMBHs or are swallowed by the SMBH, thus this is conservative in the direction that disfavor IMBHs)
\bea 
\mathcal{R} \sim \frac{\int_{0}^{\sim 10-14 {\rm Gyr}} \varphi(t) \eta(t) dt}{\int_{0}^{\sim 10-14 {\rm Gyr}} \eta(t) dt} \,. 
\eea
The sBH birth rate should not be heavily back-loaded because the lifespan of sBH forming massive progenitor stars are measured in mere millions of years. For simplicity, we can take $\eta$ to be a constant, and then it drops out from $\mathcal{R}$. 
The formation and subsequent survival rate $\varphi(t)$ is zero for $t \lesssim 1$Gyr, and is anchored at around $1/9$ at $\sim 10$Gyr by \cite{2022ApJ...929L..22R}. It likely has a peak at sometime before $10$Gyr and after $1$Gyr when IMBHs have had enough time to grow but not enough time to be captured by the SMBH yet. Conservatively, if we ignore the peak and then simply let $\varphi(t)$ be a linearly growing function between $0$ @$1$Gyr to $1/9$@$10$Gyr, and be $0$ at $<1$Gyr and $>10$Gyr, then $\mathcal{R} \approx 0.036(0.05)$ for integration out to $14(10)$Gyr. If we acknowledge the potential existence of the peak above $1/9$, but chop it off and use the area underneath it but above $1/9$ to fill in the regions of $\varphi(t)$ that fall short of $1/9$ in other parts of the $1$Gyr to $10$Gyr interval, so that $\varphi(t)$ becomes a constant there, then $\mathcal{R} \approx 0.07(0.1)$. In short, one has some grounds in arguing for an optimistic ratio of up to $\mathcal{R} \sim \mathcal{O}(10^{-1})$. So there could be as many as $\sim \mathcal{O}(10^3)$ IMBHs present in our Galactic center that are formed in the nuclear stellar cluster\footnote{On the other hand, the IMBH deposition by globular clusters falling from outside into the Galactic center is very roughly estimated to occur at a rate of $0.1-1{\rm Gyr}^{-1}$ for the Milky Way \cite{2005ApJ...618..426M,2018MNRAS.477.4423A,2023MNRAS.524.2033S}, so this channel is likely unimportant in comparison, unless the IMBHs they bring in are much larger in mass.}. A comparable number of them being absorbed by the SMBH can indeed suffice for the mass budget of the $M_{\rm SMBH} \sim 4\times 10^6\Msun$ Sgr A*. For other galaxies with larger SMBHs, the number and/or the individual masses of the IMBHs could be greater. 

\section{Adding IMBHs to the mix \label{sec:solvepc}}
\begin{figure}[tb]
  \centering
\begin{overpic}[width=0.99\columnwidth]{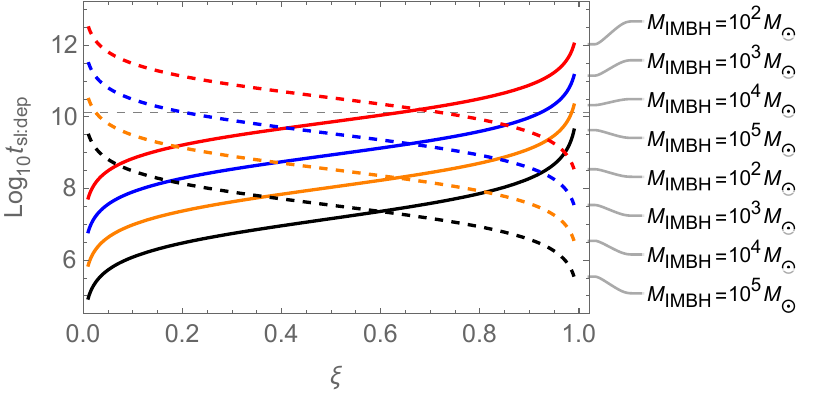}
\end{overpic}
  \caption{
The dependence of $t_{\rm sl:dep}$ on $\xi$ for a range of $M_{\rm IMBH}$ values. Solid curves are obtained using Eq.~\eqref{eq:stall}, while dashed curves are from Eq.~\eqref{eq:stall2}. The grey horizontal dashed line is the age of the Universe. 
}
	\label{fig:deptime}
\end{figure}

Now consider the scenario where effective scattering agents for the final parsec stage of the SMBH binary orbital evolution are IMBHs rather than stars. We now would have $N$ in Eq.~\eqref{eq:stall} being replaced by $N_{\rm IMBH}$ according to Eq.~\eqref{eq:BHNum}. Assume that $M_{\rm IMBH} \sim \mathcal{O}(10^5)\Msun$ as per the candidate IMBH observed in the Galactic center by \cite{2017NatAs...1..709O} (heavier IMBHs are also more likely to quickly sink into the vicinities of SMBHs \cite{2020MNRAS.496..921W}, but we explore all possibilities in Fig.~\ref{fig:deptime}), and capture efficiency of $\xi \sim 1/2$ as per discussions from Sec.~\ref{sec:IMBHpop}, then for the archetypal final parsec problem demonstration, the example shown in Fig.~1 of \cite{1980Natur.287..307B} with $M_{\rm SMBH} \sim \mathcal{O}(10^8)\Msun$, this results in $t_{\rm sl:dep} \sim \mathcal{O}(10) t_{\rm df}$ according to Eq.~\eqref{eq:stall}. 

Strictly speaking, $t_{\rm df}$ that appears as a formal segment of the $t_{\rm sl:dep}$ expression in Eq.~\eqref{eq:stall} should now be for the cluster of IMBHs, rather than stars. It only weakly depends on $N$, via a logarithm (see Eq.~\ref{eq:dftime}), and dispersion velocity should be the same as the stars (both on the order of the test mass orbital speed as determined by the cluster's gravitational potential) if virialization hasn't happened and smaller if it has progressed to some extend, while the cluster radius should be smaller due to the greater concentration of IMBHs towards the center. These changes to the parameters would further reduce $t_{\rm df}$ and subsequently $t_{\rm sl:dep}$. 
In fact, the stalling may even be further reduced in cases where the IMBH masses are towards the larger end, as the Kozai-Libov effect \cite{1962AJ.....67..591K,1962P&SS....9..719L} they excite could increase the eccentricity of the SMBH binary (see in particular Fig.~4 of \cite{2016ARA&A..54..441N} for the effects of a small mass outer perturber), which in turn causes enhanced bursts of GW during periastron passages \cite{2020arXiv200911332L} (cf., Eq.~\ref{eq:GWtime}; note this energy and angular momentum loss means we won't see the same indefinite oscillations as with the regular Kozai-Libov evolution, and the orbit tends to circularize at smaller separation), in effect expedites the takeover of the $t_{\rm gw}$ stage and thus beckons a more efficient merger. This is somewhat similar to how sBH-SMBH mergers happen. The sBHs are too light for dynamical friction to work effectively, so they get dynamically captured instead, by losing large amounts of energy when passing close to the SMBH on eccentric orbits, brought into elongation when they scatter against other objects. Besides possessing more perturbing power, the more massive IMBHs are fewer in numbers, so there may be less cancellations among their contributions to this effect, although the details of this statistics still needs to be worked out, possibly via numerical simulations. 

The impacts of these effects are more difficult to ascertain, so for our demonstration of principle, we take a conservative route and assume $t_{\rm df}$ does not change when we swap from stellar to IMBH scattering agents. Then the final parsec would take only $\sim \mathcal{O}(10^7)$yr. For thoroughness, we can also adopt the alternative expression \eqref{eq:stall2} and compare, where now $M_{\rm SMBH}=10^8\Msun$, $m_* = 10^5\Msun$, while $n_{\rm diff} \propto M_{\rm cl}^{-1}$ can be taken to be $10^{-1}$ of the stellar value (ignoring once again the impact of $\sigma_c$ to $n_{\rm diff}$; lower $\sigma_c$ due to virialization gives higher $n_{\rm diff}$ \cite{1978ApJ...226.1087C} and thus smaller $t_{\rm sl:dep}$), 
because $N_{\rm IMBH} \sim \mathcal{O}(10^{3})$ according to Eq.~\eqref{eq:BHNum} for our $\xi =1/2$ choice, which is roughly $10\%$ that of sBHs assuming Milky Way is typical in this respect, thus $10^{-4}$ of main sequence stars (see population ratios in Sec.~\ref{sec:Intro}). 
Eq.~\eqref{eq:stall2} subsequently yields $t_{\rm sl:dep} \sim \mathcal{O}(10^{7})$yr, broadly consistent with that provided by Eq.~\eqref{eq:stall} for this particular choice of $\xi$ and $M_{\rm IMBH}$ (the dependence on $\xi$ differs for Eq.~\ref{eq:stall} and Eq.~\ref{eq:stall2} though, see Fig.~\ref{fig:deptime}). 

In comparison, the whole process taking the SMBH binary separation down to the final parsec would take $\sim \mathcal{O}(10^6)$yr (note the earlier stages of dynamical friction are likely still dominated by stars or sBHs, since IMBHs congregate more towards the center, so $t_{\rm df}$ in this estimate would still be for that of the stellar population as scattering agents), while the prior merger and violent relaxation of the cores takes $\sim \mathcal{O}(10^8)$yr \cite{1980Natur.287..307B}. In other words, there is barely any stalling during the final parsec. We further plot the dependence of $t_{\rm sl:dep}$ on $\xi$ and $M_{\rm IMBH}$ in Fig.~\ref{fig:deptime}, and see that even for the smallest IMBHs that can easily form via mergers of sBHs, there are $\xi$ ranges that allow SMBH mergers to have occurred. 

We must caution though, the statistical nature of much of the underlying derivations quoted in this note would render the quantitative expressions suspect for such small IMBH population numbers. In particular, the dependence of $t_{\rm sl:dep}$ on $N$ could be weaker for smaller values of this parameter (see \cite{2004ApJ...602...93M} and references within). The IMBH population size adopted here could also be overly optimistic (e.g., we have assumed that IMBH accretion is not subdominant to SMBH mergers, when it comes to contributing to the growth of SMBH masses). Nevertheless, the qualitative heuristics should stand, and there is a three orders of magnitude headroom for further adjustments if merger within a Hubble time is all that is desired. 

\section{Discussion \label{sec:discussion}}
The key uncertainty in the proposed resolution of the final parsec problem is in the uncertainty of the IMBH population size. Observationally, the IMBHs settle in a crevice that helps them remain hidden. If they are mostly formed inside nuclear stellar clusters of massive galaxies, then they would be heavily obscured from outside view by intervening stars, dust, gas and etc. As a result, electromagnetic channel observations \cite{2020ARA&A..58..257G} would in general be rather challenging.  

As such, GW may offer an enticing addition, as these waves interact extremely weakly with matter and can thus avoid being blocked. They are also sourced directly by the holes as opposed to indirectly by environmental elements influenced by those holes, and thus could offer up more unambiguous evidences. With the GW approach, there is a tradeoff between signal strength, thus detection horizon, against population size per galaxy. This is because given a fixed sensitivity range in GW frequency, a detector will be able to see the earlier inspiral stages of binaries involving lighter holes, but the plunge and merger stages of those involving heavier holes (the frequency at merger scales inverse proportionally with the mass of the heavier hole, see e.g., \cite{2005ApJ...618..426M}). Heavier holes are rarer, and merger phases of binaries last much shorter than the inspiral phases, so per galaxy event rates are lower; but offsetting this, the mergers of binaries involving heavier holes send out stronger waves (proportional to the mass of the lighter hole in the binary \cite{2005ApJ...618..426M}), so we can detect them in further off galaxies. One should consider both source types. 

With the inspiral-focused option, ingenious schemes (e.g., \cite{2022PhRvD.105l4048S,2023MNRAS.524.2033S,2022ApJ...933..170F,2023ApJ...944...81F,2023arXiv231202983E}) have been devised to utilize the various bound configurations of SMBHs, IMBHs, and sBHs that fit into the sensitive bands of LISA \cite{2017arXiv170200786A} (see also Taiji \cite{2020ResPh..1602918L} that share similar characteristics). Assume the most optimistic scenario that the IMBH population is as high as one tenth of that of the sBHs, there should thus be very roughly one tenth (one hundredth) the number of IMBH-sBH (double IMBH) binaries as there are double sBH binaries (adopting dynamical capture during chance encounters as the leading formation mechanism and ignoring cross section and spatial distribution differences). During the operation time of LISA, there is expected to be up to $\sim \mathcal{O}(10^6)$ double sBH systems being detected in its sensitivity band \cite{2016PhRvL.116w1102S}, and after adapting for duration of stay in band according to Eq.~\eqref{eq:GWtime}, also adjusting for detection horizon, one sees that the probability of LISA detecting binaries involving IMBHs, in their inspiral stage, should not be discounted. 

Alternatively, we can opt for the merger-focused option. The most sensitive band of LISA lands at around $3\times 10^{-3}$ to $10^{-2}$Hz, so $M_{\rm SMBH} \sim \mathcal{O}(10^{6})$ and slightly lighter are the most favorable. If we restrict attention to the merger of IMBHs into SMBHs, whose final stages also fall within the LISA band, then one notes that around $\mathcal{O}(10^3)$ merger events of $\sim \mathcal{O}(10^3)\Msun$ IMBHs would form a $\sim \mathcal{O}(10^6)\Msun$ SMBH, and happening within the $\sim 10$Gyr lifetime of a galaxy would yield, on average, $\mathcal{O}(10^{-7})$ merger events per year per galaxy, or $\mathcal{O}(10^{-6})$ events per galaxy throughout the longer end estimate of LISA's operational time of $10$ years. The estimates by \cite{2004ApJ...602..603Y,2005ApJ...618..426M} is that there are $\mathcal{O}(10^8)$ SMBHs in this lower mass range within the LISA detection horizon, so the lower limit (mergers involving higher mass SMBH may also be observable if close enough) to the observable IMBH-SMBH merger events is $\mathcal{O}(10^2)$. 

Another interesting class of potential merger sources involving IMBHs are the sBH-IMBH and double IMBH mergers (as opposed to inspirals that we have already discussed). To see such events though, we would need sensitivities in the frequency range of $\sim 10^{-2}-10^1$Hz. New instruments are required, and some have been proposed that cover various parts of this range, such as Tianqin \cite{2016CQGra..33c5010L}, DECIGO \cite{2008JPhCS.122a2006K}, or proposals that utilize the Moon as a resonant-mass detector \cite{2021ApJ...910....1H,2023SCPMA..6609513L}. 

All considered, our knowledge of IMBH population size should be revolutionized in the near future, and if turns out to be too small, the final parsec problem can not possibly be resolved with the proposal laid out in this note. On the other hand, small impact parameter slingshots of IMBHs around a SMBH binary would also emit bursts of GW that are wide band and potentially detectable. Seeing such events at a sufficient copiousness could well provide direct support. 

\acknowledgements
This work is supported by the National Natural Science Foundation of China grants 12073005 and 12021003.

\appendix 
\section{IMBHs as dark matter constituent \label{sec:DM}}
Different types of galaxies exhibit markedly different rotation curves (see e.g., references in \cite{2022arXiv220914151D,2020MNRAS.499.2912F}), especially on the interior, suggesting that there may be more than one form of dark matter. Specifically, besides the physically more exotic (nonbaryonic in the nomenclature of e.g., \cite{1994ARA&A..32..531C}) types that must extend out deep beyond the stellar concentrations, there may also be more mundane astrophysical contributions (baryonic \cite{1994ARA&A..32..531C}) concentrated near the galactic centers, where baryonic matters tend to congregate, that we simply haven't been able to see. In particular, because we are seeing extended cores (manifesting as a segment of linear growth) in the rotation curves of many dark matter dominated dwarves, as well as low surface brightness and irregular galaxies \cite{1994Natur.370..629M,1994ApJ...427L...1F,2011AJ....142...24O}, for whom the astrophysical sources would more likely to be deficient, it stands to argue that the more exotic form of dark matter may perhaps assume this extended core (corresponding to constant exotic dark matter density) morphology as its universal form, while the higher density dark matter cusps needed (in fact, as we will discuss below, there isn't yet a clear necessity at the current time) to fit the cuspy (more steeply rising near center) rotation curves of some of the more baryon-abundant luminous galaxies, contain additional contributions of astrophysical origins, that concentrate at the center. 

This pragmatic \emph{a posteriori} approach (see \cite{galaxies7010027}\footnote{In this model, that half of the spacetime curvature tensor that does not need to be sourced by matter, and is the carrier of gravitational entropy as per \cite{1979grec.conf..581P}, grows from zero at the big bang to larger values later on. Its positive contribution to the overall entropy budget allows baryonic matter to come out of thermal equilibrium (which entails a decreasing contribution to entropy thus needs offsetting). As such, it is the hidden driving force behind structure formation, while baryons play a more passive role, much like dusts being swept along into the center of vortices of an unseen underlying ocean.} for an \emph{a priori} proposal for such an extended core profile; we do not need any other specific characteristics of that model though) contrasts with the main thrust of research into the traditional cold dark matter, which has the opposite problem in the galactic centers of having to wipe out the cusps that would naturally arise with cold dark matter piling up, but not seen in dwarf galaxies, by e.g., conjuring up extreme baryonic outflow events that feedback on the dark matter, towing it away from the center \cite{2022arXiv220914151D}. Theirs is in general a much harder problem, as the heat capacity of a self-gravitating system is negative, so one would need to extract enormous energy from baryonic interactions and feed it to the supposedly much heavier (in the cold dark matter scenario) dark matter sector, to prevent its collapse into steep cusps. And somehow, this energy supply must also persist, to prevent re-collapse at later times \cite{2015MNRAS.449L..90L}. In any case, looking for even more hidden masses at galactic centers beyond the SMHBs has not really been a priority. 

The astrophysical dark matter species must be difficult to observe, so they must not emit light. They must also be compact, so they do not block light from behind. As such, black holes constitute the ideal candidates. The IMBHs are the least understood, thus least constrained, so is ideal for enlistment when there is a hidden mass budget needing to be filled. Incidentally, the IMBH mass range largely overlaps with that allowed for black hole type dark matter as assessed by \cite{1994ARA&A..32..531C}, although that review does not concentrate on galactic centers. In particular, for the aforementioned cusp production, we do not need IMBHs in the extended dark matter halo, so the inadequacy of MACHOs (Massive Compact Halo Objects) as dominant dark matter components, as appraised through, e.g., microlensing techniques \cite{1995PhRvL..74.3724G}, is not pertinent.
Besides evading existing constraints, because we are trying to create an astrophysical matter cusp, i.e., increasing density towards the center, it would be natural to look to more massive compact objects that has this tendency to sink to the center through mass segregation via dynamical friction. IMBHs are natural candidates in this regard.  

It would be desirable then, to be able to estimate the typical IMBH population size that would be needed to make up for the shortfall when we move to a cored exotic dark matter profile. However, as it turns out, the detail level and accuracy of observational data is insufficient, and the only conclusion we can draw is that the IMBH contribution does not need to dominate the bulge mass. Specifically, it was noted in \cite{galaxies7010027}  that a concentrated central astrophysical mass (monopole like; incl.~potential IMBH contributions) at $\sim 4\times 10^{10} \Msun$ is sufficient for fitting to the inner segments of the M31's (closest spiral galaxy with an unobscured view to the center) rotation curve, after we migrate to an extended core exotic dark matter profile. On the other hand, the observational bulge stellar (i.e., without IMBH) mass estimates have been reported in the literature \cite{2006MNRAS.366..996G,2008MNRAS.389.1911S,2009ApJ...705.1395C,2010A&A...511A..89C,2012A&A...546A...4T} to range from $2.32\times 10^{10}\Msun$ \cite{2009ApJ...705.1395C} to as high as $6.6\times 10^{10}\Msun$ \cite{2012A&A...546A...4T}. As a result, given our current knowledge, even stellar matter alone could possibly already account for the cusp (see also \cite{1976ApJ...209..214B} for a theoretical prediction on their cuspy number density profile), without needing to invoke the sBH or SMBH black hole mass cusp in the central parsec, let alone additional contributions from IMBHs. Nevertheless, by adopting the lower limit of the stellar mass estimates, we can obtain an upper limit to the total mass of IMBHs in the bulge of M31 at $\lesssim 1.7\times 10^{10}\Msun$.

\section{Other IMBH formation channels \label{sec:formation}}
There could be many different formation channels (see e.g., \cite{2004IJMPD..13....1M,2023arXiv231112118A}), all of which should result in a large number of IMBHs if viable, some are:
\begin{itemize}
\item 
The population III \cite{2001ApJ...551L..27M} and II \cite{2017MNRAS.470.4739S} stars, containing little metal, can in theory collapse to form IMBHs at the lower end of their mass spectrum, at hundreds of solar masses. Because metallicity of early galaxies is indeed low, as per JWST observations \cite{2023NatAs.tmp..194H}, there would have been a large population of the progenitor stars, giving birth to a large population of small-end IMBHs. They would all sink to the centers of their host galaxies, where shear crowding enables mergers, thus the production of heavier IMBHs. Accretion in the gas-rich environment of the galactic centers could also spur growth of the IMBHs, where at Eddington limit, it takes only tens of mega years for an e-folding increase in mass \cite{2014ApJ...784L..38M}. 

In fact, given our current best understanding of stellar physics, there needs to be an account of where the compact remnants of the vast number of early massive stars went, and their congregation at galactic centers (they are the first generation of stars that lived very short lives, so there would be plenty of time for them to sink into the galactic centers via mass segregation), partially or wholly merging into, thus seeding the SMBHs, offers up one possibility. 

\item
The giant molecular clouds that precedes stellar clusters have total masses matching that of IMBHs \cite{2007ARA&A..45..565M}, so if there are efficient cooling processes that act in a relay fashion to collapse a significant portion of that matter into a single entity, we would end up with IMBHs reaching even the upper end of $\mathcal{O}(10^5)\text{M}_{\odot}$. Typically, as some stars get flung out of the cluster cores into the halos, taking away energy, the core collapses, until hard stellar binaries form that can supply additional energy to the cluster by scarifying (hardening) themselves. The Heggie-Hills law however suggests that generally, hard binaries tends to merge, while soft binaries gets disrupted, allowing core collapse to continue, e.g., through runaway collisions between stars, or repeated mergers of sBHs (a major hurdle to this scenario is if the merger imparts large kicks to the resulting BH, ejecting it out of the cluster; see \cite{universe6010003} however for some complications regarding estimating kick speeds), or copious transfer of matter into sBHs, all eventually leading to the plausible formation of IMBHs, as the most compact thus thermodynamically preferable terminal state. 

So far the simulations (see references in \cite{2023arXiv231112118A}) have largely been successful in cooking up IMBHs in clusters, but the percentage of success and the mass ranges are subject to physical conditions in the clusters that are difficult to ascertain, especially for those clusters residing in the earlier universe. 
We do know however, that there are thousands of stellar clusters in the Milky Way \cite{2013A&A...558A..53K}, and still more must have existed during its long evolution history which would have since dispersed by intrinsic relaxation or extrinsic galactic tidal field stripping \cite{1990ApJ...351..121C,1997ApJ...474..223G}. Although the relevant physics remain uncertain, it is not impossible that a non-negligible subset of them can manufacture IMBHs (see e.g., \cite{1975Natur.256...23B} for observational support of IMBHs inside stellar clusters). The clusters can subsequently sink into the nucleus of the host galaxies \cite{2001ApJ...562L..19E,2006ApJ...641..319P,2017MNRAS.464.3060A} by dynamical friction, delivering their IMBHs to the galactic core.  

\item 
The masses of the massive black holes in a galaxy correlates with the host's mass and velocity dispersion, so those massive black holes in dwarf galaxies may actually be IMBHs when classified according to mass (as opposed to formation history, e.g., via direct collapse of low metallicity gas clouds \cite{2003ApJ...596...34B}). As large galaxies have supposedly swallowed many dwarf galaxies as they grow, these IMBHs should sink to the center of the merged large galaxy and find their way towards the SMBH there via mass segregation \cite{2019MNRAS.482.2913B,2020MNRAS.498.2219V}. Because dwarf galaxies are numerous and close to the larger galaxies, such mergers are frequent, and so there could be a number of such IMBHs end up surrounding the SMBH. 

\end{itemize}

\newpage
\bibliography{IMBH.bbl}

\end{document}